\documentclass   {article}
\begin{document}
\renewcommand{\theequation}{\thesection.\arabic{equation}}
\def\sectioneq{{\setcounter{equation}{0}}\section}
\title{%
Doubly special quantum and statistical mechanics from quantum $\kappa$-Poincar\'e algebra}
\author{ J.\ Kowalski--Glikman\thanks{e-mail
address jurekk@ift.uni.wroc.pl}~~\thanks{Partially supported by the   KBN grant 5PO3B05620}\\ Institute for Theoretical
Physics\\ University of Wroc\l{}aw\\ Pl.\ Maxa Borna 9\\
Pl--50-204 Wroc\l{}aw, Poland} \maketitle
\begin{abstract}
Recently Amelino--Camelia proposed a ``Doubly Special Relativity'' theory with two observer independent scales (of speed and mass) that could replace the standard Special Relativity at energies close to the Planck scale. Such a theory might be a starting point in construction of quantum theory of space-time. In this paper we investigate the quantum and statistical mechanical consequences of such a proposal. We construct the generalized Newton--Wigner operator and find relations between energy/momentum and frequency/wavevector for position eigenstates of this operator. These relations indicate the existence of a minimum length scale. Next we analyze the statistical mechanics of the corresponding systems. We find that depending on the value of  a parameter defining the canonical commutational algebra one has to do either with system with maximal possible temperature or with the one, which in the high temperature limit becomes discrete.
\end{abstract}
\clearpage

\sectioneq{Introduction}

One of the cornerstones of the XXth century revolutions in
physics, which opened an opportunity to construct both
relativistic theory of gravity and quantum field theory was the
Special Theory of Relativity. It is  therefore tempting to suggest
that the a good starting point to construct the quantum
gravity might be a proper modification of Special Relativity. An
attempt to construct such a theory has been undertaken by Giovanni
Amelino-Camelia in two seminal papers \cite{gac1}, \cite{gac2}
(for a up-to-date review see \cite{gacnew}) and further
developed in \cite{jkgminl}, \cite{jkgphsp}, and \cite{rbgacjkg}.
The central idea of this approach is to introduce a  second,
observer independent scale, playing the same role as the speed of
light does in the standard special relativity. 
There are some experimental indication, resulting from the
analysis of the cosmic rays anomalies   that this scale is of order of the Planck mass (or Planck length) \cite{gacpir}.

It should be stressed  that the aim  of the programme
of construction of this ``Doubly Special Relativity'' (as the
theory was dubbed by Amelino-Camelia) is to take a class of {\em
kinematical} symmetries, being an exact counterpart of the role
played by Poincar\'{e} symmetry in Special Relativity, as a starting
point of the more ambitious attempt to understand quantum gravity.
There is therefore an important conceptual difference between this
proposal and works
in which one tries to {\em dynamically} break the Lorentz symmetry
\cite{jac}, \cite{kost}. More detailed discussion of this issue
can be found in the paper \cite{gacjkgfut}.

Technically one of the possible realizations of ``Doubly
Special Relativity'' is the so called $\kappa$-Poincar\'{e} quantum
algebra, being a quantum deformation of the standard Poincar\'{e}
algebra, derived and analyzed in \cite{lunoruto}, \cite{maru},
\cite{luruto}, \cite{luruza}, \cite{luno}. This algebra contains a
mass scale $\kappa$ which was shown in \cite{jkgminl} and
\cite{rbgacjkg} to be observer independent (of
course, the observer independent speed of light is still present
in the construction and this is the reason for the term ``doubly
special''.) It is worth mentioning that this realization leads directly to unrevealing a  non-commutative structure of space-time.
\newline

In this paper  I would like to extend the results presented in \cite{jkgphsp},   taking the Doubly Special
Relativity as a starting point. In the next section I will briefly
recall the quantum algebraic structure of the $\kappa$-Poincar\'{e}
theory and its extension to the whole of the phase space of the one
particle-system. In section 3, following the footsteps of Maggiore
\cite{mag1} I will derive the form of generalized Newton--Wigner  position
operators and try to investigate its physical properties. 
 Section 4 will be devoted to 
statistical mechanics of the $\kappa$-Poincar\'e systems.  The last
section  contains conclusions and some speculations.

\sectioneq{Preliminaries}

The  $\kappa$-Poincar\'{e} algebra,  being a quantum algebra is not
a Lie algebra, i.e., the right hand side of its commutators is not
linear in generators. In its definition there is therefore much
more freedom than in the standard Lie algebra case, since it is
possible to replace generators by any one-to-one (differentiable)
functions of them (in the Lie algebra case one can only take
linear combinations of the generators). This freedom is however
severely restricted if one takes into account the postulate that
the algebra should be (as an algebra of space-time transformation)
integrable to a group and that there should be a reasonable balance
between the algebraic and co-algebraic sector (the latter is in
one-to-one correspondence with the anti-commuting structure of
space-time.) If one takes this requirements into account the only
known candidate for the realization of the $\kappa$-Poincar\'{e}
algebra is the so-called bicrossproduct basis \cite{maru},
\cite{luruza} (we use the units in which the universal speed of light $c$ and the Planck constant equal $1$)\footnote{It should be noted at this point that the algebra used by Maggiore \cite{mag1} and recently by \cite{KaRa}, who addressed problems similar to the ones investigated in this paper,  are based on another representation of $\kappa$-Poincar\'e algebra, which, in particular does not integrate to a group, and therefore, it seems, is not physically relevant.}:

\begin{equation}\label{2.1}
  [M_{\mu\nu},M_{\rho\tau}]
= i
\left(\eta_{\mu\tau}M_{\nu\rho} -
\eta_{\nu\rho}M_{\nu\tau} +\eta_{\nu\rho}M_{\mu\tau}
  -\  \eta_{\nu\tau}M_{\mu\rho} \right),
\end{equation}

\begin{equation}\label{2.2}
 [M_{i},P_{j}] = i \epsilon_{ijk}P_{k}\, ,
\quad [M_{i},P_{0}]= 0\, ,
\end{equation}

\begin{equation}\label{2.3}
  [N_{i}, P_{j}] = i \delta_{ij}
 \left( {\kappa\over 2} \left(
 1 -e^{-2{P_{0}/ \kappa}}
\right) + {1\over 2\kappa} \vec{P}\, ^{ 2}\, \right)
- \ {1\over \kappa} P_{i}P_{j} ,
\end{equation}
\begin{equation}\label{2.4}
 \left[N_{i},P_{0}\right] = i P_{i}\, ,
\end{equation}
\begin{equation}\label{2.5}
 [P_{\mu},P_{\nu}] =  0\, ,
\end{equation}
where $P_\mu =(P_i, P_0)$ are space and time components of
four-momentum, and $M_{\mu\nu}$ are deformed Lorentz generators\footnote{Note that the term
``deformed''  describes the action of boosts on momenta and not 
 deformation of Lorentz algebra.}
consisting of rotations $M_k = \frac12\epsilon_{ijk}M_{ij} $ and
boosts $N_i = M_{0i}$, and $\kappa$ is a scale parameter.  The
Casimir operator of this algebra has the form
\begin{equation}\label{2.6}
  {\cal M}^2 = \left(2\kappa \sinh\left(\frac{P_0}{2\kappa}\right)\right)^2 -
  \vec{P}\,{}^2 e^{P_0/\kappa}.
\end{equation}

Eqs.~(\ref{2.1}--\ref{2.6}) describe the  algebraic structure of
the $\kappa$-Poincar\'{e} algebra. This algebra is, however, a quantum algebra by
construction, so it possesses additional
sectors. This so-called co-algebra sector is in a direct relation
with the non-commutativity of space-time. More precisely, if one
introduces the space-time position operators $X_\mu$ (not to be
confused with the Newton--Wigner position operators in {\em
space}, ${\cal X}^{\kappa}_i$, to be introduced in the next
section), they are to satisfy the following commutational relation
\begin{equation}\label{2.7}
  [ X_0, X_i] = -i\frac1\kappa \, X_i,
\end{equation}
Starting from this relation one can construct the class of generalized canonical commutational relations as follows. We assume that
\begin{equation}\label{2.8}
  [ X_0, P_0] = i,
\end{equation}
and
\begin{equation}\label{2.9}
  [ X_i, P_j] = -i\delta_{ij} \Phi(P_0),
\end{equation}
\begin{equation}\label{2.9a}
  [ X_0, P_i] = i \Psi(P_0) P_i,
\end{equation}
so that the rotational covariance is manifestly present, and all other commutators vanish. The Jacobi identity for the algebra (\ref{2.8} -- \ref{2.9a}) leads to the relation
\begin{equation}\label{2.10}
 \Psi(P_0) = \frac{\Phi'(P_0)}{\Phi(P_0)} + \frac1\kappa,
\end{equation}
where, of course, $\Psi(P_0) \rightarrow 0$ and $\Phi(P_0)\rightarrow1$ when $\kappa\rightarrow\infty$. 

There are two natural solutions of this relation. The one with $\Psi(P_0) = 0$, $\Phi(P_0) =e^{-P_0/\kappa}$ was analyzed in \cite{jkgphsp} and leads, in particular, to space-time equipped with Minkowski metric and the {\em physical} speed of light (massless modes) equal $1$ . The second, with $\Phi(P_0) =1$, $\Psi(P_0) = 1/\kappa$ is known in the literature as ``Heisenberg double algebra'' and seems to be more natural from the non-commutative geometry pint of view. In this case, however the space-time picture (as opposed to energy/momentum space) is not well understood. Therefore,for the sake of completeness  in this paper I will study the one-parameter class of algebras  with $ \Phi(P_0) = e^{a P_0/\kappa}$, where $a$ is  real.

\sectioneq{The generalized Newton--Wigner position operator}

In this section we will derive, following the footsteps of
Maggiore \cite{mag1}, the position operator and its commutator
algebra. The idea is to construct an operator  generalizing the Newton--Wigner relativistic position
operator \cite{NW}.

The first step in the construction is to define a scalar product
invariant under action of $\kappa$-Poincar\'{e} algebra. Here one
must be a bit careful, because, contrary to the standard case as
well as a case considered by Maggiore in \cite{mag1}, the measure
$d^4 P$ is {\em not} invariant under boost transformations
(\ref{2.3}), due to the presence of the  off-diagonal terms in
these transformations. Indeed, if we take
$$\delta P_i \equiv \varepsilon^j [N_{j}, P_{i}] = \varepsilon_i
 \left( {\kappa\over 2} \left(
 1 -e^{-2{P_{0}/ \kappa}}
\right) + {1\over 2\kappa} \vec{P}\, ^{ 2}\, \right)
- \ {1\over \kappa}\varepsilon^j P_{j}P_{i} ,
$$ $$
\delta P_0 \equiv \varepsilon^i\left[N_{i},P_{0}\right] =
\varepsilon^i P_{i}$$ with  infinitesimal $\varepsilon$ and make
use of the fact that in $n$ space dimensions $$\delta\, (d^{n+1} P)  = \sum_1^{1+n}
\frac{\partial\delta P_\mu}{\partial P_\mu}\,\, d^{n+1}P,$$ we find
easily that in the case at hands $$\delta\, d^{n+1}P = -\frac{n \varepsilon^i
P_{i}}\kappa\, d^{n+1}P.$$  We see therefore that the correct
invariant measure is given by 
\begin{equation}\label{3.0}
 d \mu = e^{n P_0/\kappa}\, d^{n+1} P
\end{equation}

Now it is straightforward to write down the $\kappa$-Poincar\'e invariant inner product  with the help of the Casimir operator (\ref{2.6}):
$$
(\varphi,\psi) =$$ $$ \int \frac{e^{n P_0/\kappa}\, d^{n+1}P}{(2\pi)^n} \,
\theta(P_0)\, \delta\left[{\cal M}^2 + \vec{P}\,{}^2
e^{P_0/\kappa}- \left(2\kappa
\sinh\left(\frac{P_0}{2\kappa}\right)\right)^2 \right]
\varphi^*(P)\psi(P) $$

\begin{equation}\label{3.1}
= \int \frac{d^nP}{2(2\pi)^n} \frac{e^{n P_0/\kappa}}{ f(P_0)}\,
\, \varphi^*(\vec{P})\psi(\vec{P}),
\end{equation}
where
\begin{equation}\label{3.1a}
  f(P_0) =\kappa\left(1-e^{-P_0/\kappa}\right) +\frac{{\cal M }^2}{2\kappa}
\end{equation}
 and  $P_0=P_0(\vec{P})$ denotes the positive solution of eq.~(\ref{2.6}).
 One important thing should be noted at this point. Namely,
 it follows from the form of the Casimir (\ref{2.6}) that the
 range of $\vec{P}$ is limited to the region
 ${\cal D} = \left\{\vec{P}:\, \vec{P}^2\leq \kappa^2\right\}$ ($\vec{P}^2= \kappa^2$
 corresponds to $P_0 = \infty$.) We therefore define the Hilbert space of
 functions $\varphi,\psi, \ldots$ to be a space of functions of class $L^2({\cal D})$.
\newline

Now we can turn to the generalization of the Newton--Wigner position
operator.  To construct it, let us assume that
\begin{enumerate}
\item the operator ${\cal X}^{\kappa}_i$ becomes
the standard Newton--Wigner operator
$$
{\cal X}^{NW}_i \equiv -i\hbar \left(\frac{\partial}{\partial P_i} - \frac{P_i}{2P_0^2}\right)
$$
in the limit $\kappa\rightarrow\infty$;
\item  it is hermitean with respect to the inner product (\ref{3.1});
\item  the commutator $[{\cal X}^{\kappa}_i, P_j] = [X_i, P_j] = -i\delta_{ij} \Phi(P_0)$, cf.~(\ref{2.9}).
\end{enumerate}

Given that the action of rotations in the $\kappa$-Poincar\'{e}
algebra is standard it is natural therefore to postulate
\begin{equation}\label{3.2}
 {\cal X}^{\kappa}_i = -i \left(\alpha(P_0;\kappa)\frac{\partial}{\partial P_i}
 - \beta (P_0;\kappa) \frac{P_i}{2P_0^2}\right),
\end{equation}
where both $\alpha(P_0;\kappa)$ and $\beta (P_0;\kappa)$  go to
$1$ when $P_0/\kappa\rightarrow0$. In this way we satisfied the
first requirement; let us now turn to the hermicity condition.
Using the relation
$$\frac{\partial P_0}{\partial P_i} =
e^{P_0/\kappa} \frac{P_i}{\kappa \sinh \frac{P_0}\kappa -
\frac{\vec{P}^2}{2\kappa} e^{P_0/\kappa}} = e^{P_0/\kappa}
\frac{P_i}{f(P_0)}
$$
one gets
\begin{equation}\label{3.3}
  \beta(P_0;\kappa) = -P_0^2\,{e^{-(n-1)P_0/\kappa}} \,
  \frac{d\, }{d P_0}\left(\frac{ e^{nP_0/\kappa}\, \alpha(P_0;\kappa)}{  f(P_0)}\right).
\end{equation}
On the other hand the third condition gives 
\begin{equation}\label{3.3a}
 \alpha(P_0;\kappa) = \Phi(P_0)
\end{equation}
so that the generalized  
the generalized position operator is now completely defined. 

Knowing this, we can try to find the velocity of a particle. One easily finds that
\begin{equation}\label{3.4} \dot{\cal X}^{\kappa}_i \equiv i \left[ {\cal X}^{\kappa}_i, P_0\right]
= \alpha(P_0;\kappa) \frac{\partial P_0}{\partial P_i} =\Phi(P_0) \frac{e^{P_0/\kappa}\, P_i}{\kappa \sinh \frac{P_0}\kappa - \frac{\vec{P}^2}{2\kappa}
e^{P_0/\kappa}}.\end{equation}

If we now substitutes $\Phi(P_0) = e^{-P_0/\kappa}$, we find
\begin{equation}\label{3.5}
 \dot{\cal X}^{\kappa}_i = \frac{ P_i}{\kappa \sinh \frac{P_0}\kappa - \frac{\vec{P}^2}{2\kappa}
e^{P_0/\kappa}}
\end{equation}
in a complete agreement with the expression for velocity found in \cite{jkgphsp}. This result provides an important
 proof of the internal consistency of the scheme presented here.

Knowing $\alpha(P_0;\kappa)$, from (\ref{3.3}) one can easily derive the
expression for $\beta$, to wit
\begin{equation}\label{3.6}
 \beta(P_0;\kappa) = -P_0^2\, \frac{e^{P_0/\kappa}}{2f(P_0)}
 \left(\frac{n \Phi(P_0)}{\kappa } + \Phi'(P_0) -
\frac{e^{-P_0/\kappa} \Phi(P_0) }{ f(P_0)}\right).
\end{equation}
Using this expression one can readily write down the generalized Newton--Wigner position operator
$${\cal X}^{\kappa}_i = -i\left[ \Phi(P_0)\, \frac{\partial}{\partial P_i} +  \frac{e^{P_0/\kappa}\, P_i}{2f(P_0)}
 \left(\frac{n \Phi(P_0)}{\kappa } + \Phi'(P_0) -
\frac{e^{-P_0/\kappa} \Phi(P_0) }{ f(P_0)}\right)\right]
$$
and compute the commutator
\begin{equation}\label{3.7}
  \left[{\cal X}^{\kappa}_i, {\cal X}^{\kappa}_j\right] = -i \frac{\Phi'(P_0)\, e^{P_0/\kappa}}{ f(P_0)} \, M_{ij},
\end{equation}
where $$M_{ij} = i \left( P_i \frac{\partial}{\partial P_j} - P_j \frac{\partial}{\partial P_i}\right)$$
is a generator of rotations. This result is of course similar to
the result of Maggiore \cite{mag1}, simply because the commutator (\ref{3.7}) must
be proportional to angular momentum in order to satisfy, along with (\ref{3.5}) the Jacobi
identity. It is worth noticing that we have to do with commuting space if and only if $\Phi(P_0) =1$.
\newline

Knowing the formula for Newton--Wigner position operator we can address the question concerning the relation between momentum and wavelength (and energy and frequency). To do that let us consider an equation for position eigenstate in the  one dimensional massless case with $\Phi(P_0) = e^{aP_0/\kappa}$. By making use of the identity $ e^{-P_0/\kappa}{\partial}/{\partial P}={\partial}/{\partial P_0}$ and $P = f(P_0)$ (which hold for ${\cal M} =0$) it is handy to write the eigenfunctions as functions of $P_0$, to wit
$$
  \xi \Psi_\xi(P_0) = {\cal X}^{\kappa}\Psi_\xi(P_0) 
=$$
\begin{equation}\label{3.8}= -i\left[ e^{(a+1)P_0/\kappa}\, \frac{\partial}{\partial P_0} +  \frac{e^{(1+a)P_0/\kappa}}{2}
 \left(\frac{1+a}{\kappa } -
\frac{e^{-P_0/\kappa} }{ f(P_0)}\right)\right]\Psi_\xi(P_0),
\end{equation}
and whose solution is
\begin{equation}\label{3.9}
\Psi_\xi(P_0) = \exp\left[\frac{i\xi\kappa}{(a+1)}\left(1-  e^{-(a+1)P_0/\kappa}\right)\right]\, \psi(P_0),
\end{equation}
where $\psi(P_0)$ is a factor which will be irrelevant for the discussion to follow. In (\ref{3.9}) we  adjusted the integration constant so as to assure that the result becomes the standard one in the limit $P_0/\kappa\rightarrow0$.

From (\ref{3.9}) one readily finds the relation between momentum and the wavevector $k = 2\pi/\lambda$, where $\lambda$ is the wavelength (re-introducing $\hbar$).
\begin{equation}\label{3.11}
 \hbar k =   \frac{\kappa}{a+1}\left(1-  e^{-(a+1)P_0/\kappa}\right) = \frac{\kappa}{a+1}\left( 1 - \left(1 - \frac P\kappa\right)^{a+1}\right)
\end{equation}
To find the relation between frequency $\omega$ and energy $P_0$, we postulate that the velocity (\ref{3.5}) equals group velocity:
\begin{equation}\label{3.12}
 \frac{d\omega(P_0)}{dk(P_0)} = e^{(a+1) P_0/\kappa} ,
\end{equation}
from which it follows immediately that 
\begin{equation}\label{3.13}
 P_0 = \hbar\omega.
\end{equation}
Of course in the limit $P_0/\kappa\rightarrow0$ relations (\ref{3.11}), (\ref{3.13}) become the standard ones. Observe that in the case $a=-1$, $k=\omega$,  which reflects the fact that for $a=-1$ the space-time metric is Minkowskian, cf.~\cite{jkgphsp}. For $a <-1$ it follows from (\ref{3.11}) that the values of the wavevector belong to the interval $k \in [0, \kappa/\hbar(a+1))$ which means that there exists a minimal value of the wavelength 
\begin{equation}\label{3.14}
  \lambda_{min} = \frac{2\pi\hbar(a+1)}{\kappa}.
\end{equation}

\sectioneq{Statistical mechanics}

In this section I will study the statistical mechanics of
systems described in the preceding section. It should be noted that similar
studies has been performed in \cite{stjo} and \cite{KaRa}, however the methods used in these papers differ from the one to be presented below. 
\newline

To start with, consider the partition function\footnote{Since we will be
interested mostly in the high temperature regime, to make
computations simpler and more transparent we will consider the
massless case only, ${\cal M}=0$.}
\begin{equation}\label{4.1}
 Z = \sum e^{- \beta P_0/kT},
\end{equation}
where $\beta = 1/kT$. The partition function merely expressed the
fact that the system  in consideration is in thermal equilibrium
with large thermal bath, so it does not depend whatsoever on
dispersion relation. However the sum is taken over elementary cells in the phase space, which means that the measure of the corresponding integral in the large volume limit must be a descendant of the  phase space Liouville measure. In the standard case this amounts of introducing the multiplicative factor $\hbar^{-3}$ (in $3$ dimensions); in the case at hands it follows that the right factor is\footnote{To  find the Liouville measure one computes the determinant of the symplectic form, i.e., the determinant of the matrix of Poisson brackets of phase space variables and multiplies it by an appropriate power of $\hbar$. It is easy to see that the $[X, X]$ commutator does not contribute to this determinant.} $\hbar^{-3} e^{-3aP_0/\kappa}$. 
Given that, one can readily apply the methods
of textbook statistical mechanics to compute the internal energy
of the ``deformed'' gas of spinless bosons in volume $V$. In the first step we  count the number of states with wavevectors between $k$ and $k+dk$. According to what was said above it equals
\begin{equation}\label{4.2}
 \frac{V}{(2\pi)^3\hbar^3} \,e^{-3aP_0/\kappa}\, 4\pi k^2 dk.
\end{equation}
It is convenient to express this measure in terms of energy $P_0$. Using (\ref{3.11}) one gets 
\begin{equation}\label{4.3}
 d\mu = \frac{V\kappa^2}{2\pi^2\hbar^3 (a+1)^2}\, \left(1- e^{-(a+1) P_0/\kappa}\right)^2\, e^{-(4a+1)P_0/\kappa}\, dP_0.
\end{equation}
Then it follows immediately that the internal energy equals 
$$
U = \int_0^\infty d\mu \frac{P_0}{e^{\beta P_0} -1}.
$$

Let us first compute this integral for $a=-1$. Taking the limit, we find
$$
  U = \frac{V}{2\pi^2\hbar^3}\int_0^\infty dP_0\,    e^{3P_0/\kappa} \,  \frac{P_0^3}{e^{\beta P_0} -1},
$$
which can be easily evaluated in terms of  the third
derivative of the digamma function $\psi$:
\begin{equation}\label{4.4}
  U = \frac{V }{2\pi^2\hbar^3} \,k^4 T^4 \psi^{(3)}\left(1 - \frac{3kT}{\kappa} \right) ,
\end{equation}
where we replaced $\beta$ with $1/kT$. For low temperatures ($kT/\kappa \ll1$) this formula reproduces, of course,  the standard  result $U \sim T^4$ up to the terms of order $O((kT)^5/\kappa)$. It is   also clear that the internal energy (\ref{4.4}) diverges for temperatures $kT> \kappa/3$. We see therefore that in this case we have to do with limiting (Hagedorn) temperature.

To find an expression for pressure ${\cal P}$ we make use of the second law of thermodynamics which can be rephrased in the form of the differential equation
\begin{equation}\label{4.5}
 u \equiv \frac{U}{V} = T \frac{d{\cal P}}{dT} - {\cal P},
\end{equation}
from which and (\ref{4.4}) it follows immediately that
$$
  {\cal P} = -\frac{1}{2\pi^2\hbar^3}\,  \left[\frac\kappa{3}\,(kT)^3\,\psi^{(2)}\left(1 - \frac{3kT}{\kappa} \right) + \frac{2\kappa^2}{9} \,(kT)^2\,\psi'\left(1 - \frac{3kT}{\kappa}\right) + \right.$$ \begin{equation}\label{4.6}\left. +\frac{2\kappa^3}{27}\,(kT)\,\psi\left(1- \frac{3kT}{\kappa}\right) \right]+ \frac{2\gamma \kappa^3 kT}{27} ,
\end{equation}
where the last term, corresponding to the integration constant has been fixed so that in the limit $T\rightarrow0$, $u = 3P$, and $\gamma$ is the Euler gamma constant. Again, for $k T\rightarrow \kappa/3$, the pressure diverges.
\newline

Let us now turn to the case $a>-1$. From the form of the integral it is clear that the internal energy would behave qualitatively similarly to the case considered above (there is a maximal temperature) if $a< -1/4$. Indeed we find 
$$
  U = \frac{V\kappa^2}{2\pi^2\hbar^3 (a+1)^2} \, (kT)^2 \left[\psi'\left(1 + \frac{(1+4a)kT}{\kappa} \right) -2 \psi'\left(1 + \frac{(2+5a)kT}{\kappa}\right) + \right.$$ \begin{equation}\label{4.7} \left. +\psi'\left(1+\frac{(3+6a)kT}{\kappa}\right)\right]
\end{equation}
Again, by making use of (\ref{4.5}) we find an expression for pressure
$$
  {\cal P} = \frac{V\kappa^3\, kT}{2\pi^2\hbar^3 (a+1)^2}  \left[\frac1{1+4a}\,\psi\left(1 + \frac{(1+4a)kT}{\kappa} \right) -\frac{2}{2+5a} \psi\left(1 + \frac{(2+5a)kT}{\kappa}\right)  \right.$$ \begin{equation}\label{4.8}\left. +\frac1{3(1+2a)}\psi\left(1+\frac{3(1+2a)kT}{\kappa}\right) + \frac{2\gamma(1+a)^2}{3 (1 + 2a) (1 + 4 a) (2 + 5 a)} \right].
\end{equation}

We see therefore that for $a \geq -1/4$ there is no maximal temperature and in the hogh temperature limit the internal energy behaves as
\begin{equation}\label{4.9}
  U \sim {V\kappa^3}\, kT.
\end{equation}
This result is of extreme interest if one recalls  that in $d$
space dimension internal energy for the standard gas of bosons
behaves as $\sim T^{d+1}$. One could therefore interpret
(\ref{4.9}) as the fact that at high temperature the system
becomes zero-dimensional, in other words it possesses only finite
number of degrees of freedom. It is worth recalling at this point the formula
 for internal energy of a crystal consisting of
$N$ atoms at high temperature limit:
\begin{equation}\label{4.10}
 U_{cr} = U_0 + 3 N kT,
\end{equation}
where $U_0$ is the zero point energy. Comparing these two formulas
we see that one can identify
\begin{equation}\label{4.11}
 N \sim {V\kappa^3}
\end{equation}
which, since $\kappa$ is of order of the inverse of the Planck length,  indicates
that at high temperatures we have to do with a system whose
degrees of freedom are confined in elementary cells of Planck volume
each. 
 This implies in turn that at high temperatures the Hilbert space
 of the system becomes finite dimensional.

\sectioneq{Conclusions and outlooks}

Let us summarize the major result of this paper.

First of all we constructed the generalized Newton--Wigner position operators which can be treated as a first step in investigations of quantum mechanical properties of the systems with two observer independent scales. It turns out that, except for the case $a=0$, these position operators do not commute, which reflects the noncommutative structure of space-time we started with. Using these operators we constructed position eigenstates, and comparing them with the analogous states of the standard quantum mechanics we  found relation between energy/momentum and frequency/wavevector that indicate the existence of a minimal length. Finally we investigated the statistical mechanics of doubly special systems, discovering that there are two possible high temperature regimes: one with maximal temperature and the second in which the system becomes effectively zero-dimensional.

It would be of extreme interest to further investigate the physical properties of the systems described in this paper. In particular, it would be very interesting to understand the physical nature of maximal temperature effect for $a\leq -1/4$, namely is it ``real'', or merely expresses a limitation of thermodynamical limit. On the other hand, the results in the case $a\geq -1/4$ seem to support the naive expectation that at ultra-high energy regime we have to do with one degree of freedom per Planck volume. A question arises as to if this result can be reconciled with the holographic principle \cite{ng}, which is regarded by many to be a fundamental principle of Planck scale physics.  These questions will be addressed in a future publication.

\end{document}